\documentclass[11pt,a4paper]{article}

\usepackage{jheppub}

\usepackage{color}
\usepackage{colordvi}
\usepackage{changes}

\usepackage{epsfig,epsf}
\usepackage{amsmath}
\usepackage{amsthm}
\usepackage{amsfonts}
\usepackage{amssymb}
\usepackage{dsfont}
\usepackage{epstopdf}
\usepackage{multirow}

\usepackage{rotating}

\usepackage{marvosym}

\usepackage{array}   
\newcolumntype{C}{>{$}c<{$}}

\usepackage{slashed}

\usepackage[active]{srcltx}

\def\II{\hbox{{1}\kern-.25em\hbox{l}}}

\def\II{\hbox{{1}\kern-.25em\hbox{l}}}

\title{
Kinematic twist--three contributions to pseudo- and quasi-GPDs and translation invariance}

\author{V. M. Braun}

\affiliation{
   Institut f\"ur Theoretische Physik, Universit\"at
   Regensburg, D-93040 Regensburg, Germany}

\emailAdd{vladimir.braun@ur.de}

\abstract{
We present explicit expressions for the tree-level ``kinematic'' twist-three contributions to the nucleon matrix elements
of gauge-invariant nonlocal quark-antiquark operators which can be used in lattice calculations
of generalized parton distributions (GPDs). These contributions in particular restore the 
translation invariance  of the results up to higher twist four.
The calculated twist-three corrections are logarithmically enhanced as compared 
to the leading twist, and are discontinuous at the kinematic points $x=\pm\xi$.  
       }


\begin{document}
\maketitle

\section{Introduction}\label{sec:intro}

A very high luminosity of the JLAB 12 GeV accelerator~\cite{Dudek:2012vr} and, in future,  
the Electron Ion Collider (EIC)~\cite{Accardi:2012qut} will allow one to 
study hard exclusive and semi-inclusive reactions with identified particles in the final state with 
unprecedented precision. A major goal of this ambitious research program is to 
understand the full three-dimensional proton structure. In particular generalized parton 
distributions (GPDs)~\cite{Mueller:1998fv, Diehl:2003ny,Belitsky:2005qn} emerge as an important research object which 
incorporates the information on the transverse distance separation of quark and gluons with fixed 
momentum fractions.
Deeply-virtual Compton scattering (DVCS) \cite{Ji:1996nm,Radyushkin:1996nd} is generally accepted as the
``gold-plated'' process that would have the highest potential impact for the transverse distance imaging.
The main challenge of these studies is that GPDs are functions of three variables (not counting the scale dependence). 
Their extraction requires massive amount of data and very high precision for both experimental and theory inputs.  

It is therefore generally accepted that any additional information on GPDs from lattice calculations 
would be extremely important and should be used, e.g.~\cite{Riberdy:2023awf}, in global fits in combination with 
the experimental data.
Lattice calculations of the lowest moments of GPDs defined through matrix elements of local composite operators 
have been performed for a long time already and are gaining maturity~\cite{Bali:2018zgl,Alexandrou:2019ali,Alexandrou:2022dtc}.
The second moments are related to the gravitational form factors of the proton which are currently receiving 
a lot of attention, see \cite{Burkert:2023wzr} for a recent review.
An alternative approach, originally suggested in \cite{Ji:2013dva} for PDFs, is to calculate on the lattice 
nucleon (hadron) matrix elements of gauge-invariant nonlocal operators, usually referred 
to as ``quasi'' or ``pseudo'' distributions. 
For large hadron momenta they can be related to GPDs using light-ray operator product expansion or, equivalently, collinear factorization.
A general review of this technique and further references can be found in~\cite{Radyushkin:2019mye,Ji:2020ect}.
 The specific application to GPDs was worked out in \cite{Ji:2015qla,Xiong:2015nua,Liu:2019urm,Radyushkin:2019owq} and the first
 proof-of-the-principle lattice calculations of GPDs in this approach have been performed recently 
\cite{Lin:2020rxa,Alexandrou:2020zbe,Lin:2021brq,Scapellato:2021uke,Bhattacharya:2022aob,Bhattacharya:2023ays}.

In this work we will be dealing with one particular issue that has been observed 
in \cite{Bhattacharya:2022aob} in lattice calculations of the GPDs in the quasi-distribution approach,
namely that the restriction to the leading twist contributions leads to violation of translation invariance.
This problem is well known from  the DVCS studies, in which case it was 
shown~\cite{Braun:2011zr,Braun:2011dg,Braun:2012bg,Braun:2012hq}
that translation invariance and also electromagnetic gauge invariance of the Compton tensor 
are restored by adding specific ``kinematic'' higher-twist corrections. Such contributions are currently known
in DVCS to twist-four accuracy for the nucleon and twist-six for scalar targets \cite{Braun:2022qly}. 
The case of quasi- (or pseudo-) distributions is in principle similar but is technically more complicated 
starting at twist-four level. In this work we demonstrate restoration of translation invariance to twist-three accuracy, corresponding to
taking into account contributions that are generically suppressed by one power of the large momentum.  
Our expressions for these corrections agree, and in fact can be inferred from the existing calculations of the twist-three kinematic 
corrections in DVCS~\cite{Belitsky:2000vx,Kivel:2000rb,Kivel:2000cn}. For scalar targets we obtain
\begin{align}
 \langle p'|\bar q(z_1v)\gamma_\mu q(z_2 v) |p\rangle &= \int_{-1}^1\!\!dx\,  e^{- i (Pv) [z_1(\xi-x)+z_2(x+\xi)]}
\biggl\{
2 P_\mu H(x,\xi) +
  \Delta^\perp_\mu G(x,\xi)
\biggr\} +\ldots
\notag\\
G(x,\xi)\Big|_{x>\xi}&= 
- \int_{x}^1 \!\!\frac{dy}{y^2-\xi^2} \big(y \partial_\xi + \xi\partial_y\big) H(y,\xi)\,, 
\notag\\
G(x,\xi)\Big|_{-\xi<x<\xi}&= 
\frac{1}{2} \int_{-1}^x \!
\frac{dy}{y-\xi} 
\left(\partial_\xi  + \partial_y \right) H(y,\xi) 
- \frac{1}{2} \int_{x}^1 \!
\frac{dy}{y+\xi}  \left(\partial_\xi  - \partial_y \right) H(y,\xi)\,, 
\notag\\
G(x,\xi)\Big|_{x<-\xi}&= 
+ \int_{-1}^x \!\frac{dy}{y^2-\xi^2} \big(y \partial_\xi + \xi\partial_y\big) H(y,\xi)\,, 
\label{result:scalar}
\end{align}
where $\partial_\xi = \partial/\partial\xi$,  $\partial_x = \partial/\partial x$, 
$v^\mu$ is a given four-vector (spacelike or timelike), $z_1$ and $z_2$ are arbitrary (real) numbers,
\begin{align}
 P_\mu = \frac12 (p+p')_\mu \,, \qquad \Delta_\mu = (p'-p)_\mu\,, \qquad \xi = \frac{(pv)- (p'v)}{(pv)+ (p'v)} 
= -\frac12 \frac{(\Delta v)}{(Pv)}, 
\label{notations1}
\end{align} 
and 
\begin{align}
 \Delta_\mu^\perp &= \Delta_\mu \!+\!2\xi P_\mu\,, \qquad (v\cdot \Delta) = 0\,.
\end{align} 
The Wilson line connecting the quark and the antiquark along the $v_\mu$ direction is implied.
The expression in \eqref{result:scalar} is invariant under translations along the line connecting the quark and the 
antiquark
\begin{align}
 \langle p'|\bar q((z_1+\delta)v)\gamma_\mu q((z_2+\delta) v) |p\rangle &=
e^{i(\Delta v) \delta}\langle p'|\bar q(z_1v)\gamma_\mu q(z_2 v) |p\rangle 
\end{align}
up to twist-four corrections (shown by ellipses) that are generally down by two powers of the momentum.
 
The ``kinematic'' twist-three GPD $G(x,\xi)$ is enhanced logarithmically at $x\to \pm \xi$ and is discontinuous 
at these points (see also~\cite{Belitsky:2000vx}):
\begin{align}
 G(\xi+\epsilon,\xi) - G(\xi-\epsilon,\xi)
&=   - \frac{1}{2}\,\partial_\xi\, \text{PV} \int_{-1}^1 \!\frac{dy}{y-\xi} H(y,\xi)\,.  
\label{disc}
\end{align}
The corresponding expressions for a spin-1/2 target (nucleon) have similar structure but are more cumbersome due to proliferation of 
Dirac structures. They will be given in the text.
Our suggestion is that such kinematic contributions have to be either taken into account explicitly in lattice calculations
of GPDs in the pseudo- or quasi-PDF approach, 
or removed by taking suitable projections in Lorentz/Dirac indices. Otherwise, the difference in the results 
obtained using symmetric and asymmetric frames would need to be considered as an intrinsic uncertainty of the calculation, cf.
a detailed discussion in~\cite{Guo:2021gru} for DVCS.

\section{Light-ray operator product expansion}\label{sec:OPE}
Consider
\begin{align}
  \mathcal{O}^{(\gamma_\mu)}(z_1,z_2) &= \bar q(z_1v)\gamma_\mu [z_1v,z_2v] q(z_2 v)\,, 
\label{O}
\end{align}
where
\begin{align}
[z_1v,z_2v] &= \exp\left\{ig \int_{z_2}^{z_1}\!\!du\, v_\mu A^\mu(uv) \right\},
\label{WL}
\end{align}
which we {\it define} as a generating function (formal Tailor expansion) for the (renormalized) local operators.
\footnote{
The operators \eqref{O} are, by definition, analytic functions of the quark-antiquark separation.  
At tree level, their matrix elements correspond to pseudo- or quasi-distributions. 
Beyond tree level, matrix elements of these operators are multiplied by 
perturbatively calculable coefficient functions that contain all singularities at $v^2\to 0$, see 
\cite{Balitsky:1990ck} for details.
} 

The operators $\mathcal{O}^{(\gamma_\mu)}(z_1,z_2)$ can be expanded in contributions of different twist~\cite{Balitsky:1987bk}:
\begin{align}
 \mathcal{O}^{(\gamma_\mu)} &=  [\mathcal{O}^{(\gamma_\mu)}]^{t2} +  [\mathcal{O}^{(\gamma_\mu)}]^{t3} +
 [\mathcal{O}^{(\gamma_\mu)}]^{t4}  +\ldots\,,
\end{align}
--- the light-ray operator product expansion. 

To explain this construction, consider symmetric points $z_1=-z_2 =z$ for simplicity. By definition
\begin{align}
 \mathcal{O}^{(\gamma_\mu)}(z,-z) &= \sum_n \frac{z^n}{n!} v_{\mu_1}\ldots v_{\mu_n} 
\bar q(0)\! \stackrel{\leftrightarrow}{D}_{\mu_1} \ldots \! \stackrel{\leftrightarrow}{D}_{\mu_n} \gamma_\mu q(0)\,.
\end{align}
The local operators on the r.h.s. of this equation are neither symmetric in all indices, nor traceless, 
and
in order to separate the
leading twist contribution we need to do the symmetrization and trace subtraction explicitly. The idea is to 
assemble this series back into a nonlocal expression. In particular the symmetrization over Lorentz indices 
can be performed as~\cite{Balitsky:1987bk}  
\begin{align}
&  [\mathcal{O}^{(\gamma_\mu)}(z,-z)]^{sym} =
\notag\\
=& \sum_n \frac{z^n}{n!} v_{\mu_1}\ldots v_{\mu_n} 
\biggl\{
\frac{1}{n\!+\!1}\bar q(0)\! \stackrel{\leftrightarrow}{D}_{\mu_1} \ldots \! \stackrel{\leftrightarrow}{D}_{\mu_n} \gamma_\mu q(0) 
+
\frac{n}{n\!+\!1}\bar q(0)\! \stackrel{\leftrightarrow}{D}_{\mu} \stackrel{\leftrightarrow}{D}_{\mu_1} \ldots \! \stackrel{\leftrightarrow}{D}_{\mu_{n-1}} \gamma_{\mu_n} q(0)\biggr\}
\notag\\=&~
\frac{\partial}{\partial v^\mu} \int_0^1\!du\, \mathcal{O}^{(\slashed{v})}(uz,-uz)\,.
\end{align}
The generalization to arbitrary quark positions is trivial so that one obtains
\begin{align}
  [\mathcal{O}^{(\gamma_\mu)}(z_1,z_2)]^{t2} &= \partial_\mu \int_0^1\!du\, [\mathcal{O}^{(\slashed{v})}(uz_1,uz_2)]_{lt}\,,
\notag\\
  [\mathcal{O}^{(\gamma_\mu\gamma_5)}(z_1,z_2)]^{t2} &= \partial_\mu \int_0^1\!du\, [\mathcal{O}^{(\slashed{v}\gamma_5)}(uz_1,uz_2)]_{lt}\,,
\label{t2}
\end{align} 
where $\partial_\mu = \partial/\partial v^\mu$ and $[\ldots]_{lt}$ (the leading-twist projection) corresponds to the subtraction of traces.
Explicit expressions for the leading-twist projection operator for nonlocal operators in different representations 
can be found in \cite{Balitsky:1987bk,Braun:2011dg}. 

The (collinear) twist-three contributions are more involved. 
They include ``genuine'' twist-three contributions of quark-antiquark-gluon operators that are irrelevant 
for our present purposes,%
\footnote{Breaking of translation invariance due to genuine twist-three corrections is a twist-four effect that is beyond our accuracy.} 
and ``kinematic'' contributions of leading-twist operators. 
Retaining the latter ones only, one obtains~\cite{Braun:2011dg}
\begin{align}
&[\mathcal{O}^{(\gamma_\mu)}(z_1,z_2)]^{t3} =
\notag\\&= 
\frac12 \int_0^1\!udu \!\int_{z_2}^{z_1}\frac{dw}{z_{12}} 
\biggl\{
[(vd)\partial_\mu - (v\partial) d_\mu]
\Big(z_1 [\mathcal{O}^{(\slashed{v})}(uz_1,uw)]_{lt} + z_2 [\mathcal{O}^{(\slashed{v})}(uw,uz_2)]_{lt} \Big)
\notag\\&\quad
-i\epsilon^{\rho\nu\sigma\mu} x_\rho \partial_\sigma d_\nu  
\Big (z_1 [\mathcal{O}^{(\slashed{v}\gamma_5)}(uz_1,uw)]_{lt}
 - z_2 [\mathcal{O}^{(\slashed{v}\gamma_5)}(uw,uz_2)]_{lt}\Big)
\biggr\} + \mathcal{O}(v^\mu, v^2) + \ldots,
\notag\\
&[\mathcal{O}^{(\gamma_\mu\gamma_5)}(z_1,z_2)]^{t3} =
\notag\\&= 
\frac12 \int_0^1\!udu \!\int_{z_2}^{z_1}\frac{dw}{z_{12}} 
\biggl\{
[(vd)\partial_\mu - (v\partial) d_\mu]
\Big(z_1 [\mathcal{O}^{(\slashed{v}\gamma_5)}(uz_1,uw)]_{lt} + z_2 [\mathcal{O}^{(\slashed{v}\gamma_5)}(uw,uz_2)]_{lt} \Big)
\notag\\&\quad
-i\epsilon^{\rho\nu\sigma\mu} v_\rho \partial_\sigma d_\nu  
\Big (z_1 [\mathcal{O}^{(\slashed{v})}(uz_1,uw)]_{lt}
 - z_2 [\mathcal{O}^{(\slashed{v})}(uw,uz_2)]_{lt}\Big)
\biggr\}
+ \mathcal{O}(v^\mu, v^2) + \ldots,
\label{t3}
\end{align} 
where $\epsilon^{0123} =1$, $z_{12} = z_1-z_2$, and $d_\mu$ is the derivative over the total translation:
\begin{align}
 \langle p'| d^\mu \mathcal{O}(z_1,z_2)|p\rangle = i(p'-p)^\mu \langle p'|\mathcal{O}(z_1,z_2)|p\rangle = i\Delta^\mu \langle p'|\mathcal{O}(z_1,z_2)|p\rangle 
\,.
\end{align} 
The ellipses stand for ``genuine'' twist-three quark-antiquark-gluon contributions.

The terms $\mathcal{O}(v^\mu)$ and $\mathcal{O}(v^2)$ in \eqref{t3} (see \cite{Braun:2011dg}) 
are by themselves twist-four and ensure accurate separation between twist-three and twist-four 
corrections. They are beyond our present accuracy and will be neglected. For the same reason, in this work we
can ignore the leading twist projection operators and substitute $[\mathcal{O}(z_1,z_2)]_{lt}\mapsto \mathcal{O}(z_1,z_2)$ at all places.  
The expressions in \eqref{t2} and \eqref{t3} are valid for arbitrary matrix elements and present the starting point for our analysis.

An alternative approach could be to follow the technique adopted in Refs.~\cite{Belitsky:2000vx,Kivel:2000rb,Kivel:2000cn}, which
makes use of a set of operator identities derived in \cite{Balitsky:1987bk}. Taking off-forward matrix elements of these 
identities one obtains recurrence relations for moments that can be converted to differential equations for twist-three GPDs.  
Solving these equations with an appropriate boundary condition one obtains the result. 
The approach used in this work is much more direct.

\section{(Pseudo)scalar target}

The GPD $ H_q(x,\xi,\Delta^2)$ for a spin-zero target can be defined as the matrix element of the 
leading-twist light-ray operator
\begin{align}\label{defGPD}
\langle p^\prime|\mathcal {O}^{(\slashed{v})}(z_1,z_2)|p\rangle &= 
2 (Pv) \int_{-1}^1 dx\, e^{-i (Pv)[z_1(\xi-x)+z_2(x+\xi)]}  H(x,\xi,\Delta^2) + \mathcal{O}(v^2)\,.
\end{align}
%
As already mentioned, the leading-twist projection only affects twist-four 
contributions $\mathcal{O}(v^2)$ and can be ignored for our purposes. 
Using this definition and Eq.~\eqref{t2} we obtain for the twist-two contribution
for the operator with an open Lorentz index, see Appendix~\ref{App:t2}:
\begin{align}
\langle p^\prime|  [\mathcal{O}^{(\gamma_\mu)}(z_1,z_2)]^{t2}|p\rangle  &=
 2 P_\mu \int_{-1}^1 dx\,  e^{- i (Pv) [z_1(\xi-x)+z_2(x+\xi)]} H(x,\xi)
\notag\\ &\quad
-  \Delta^\perp_\mu  \partial_\xi \biggl\{ 
\int_0^1\!du\! \int_{-1}^1\! dx\, e^{- iu (Pv) [z_1(\xi-x)+z_2(x+\xi)]}  H(x,\xi)\biggr\},
\label{H:t2}
\end{align}
where $\partial_\xi = \partial/\partial \xi$ and 
\begin{align}
 \Delta_\perp^\mu = \Delta_\mu + 2 \xi P_\mu\,, \qquad (v\Delta_\perp) = 0\,.
\end{align}
In order to simplify notation, here and below we do not show the dependence of the GPD on the invariant momentum transfer $\Delta^2$.

The contribution in the second line in \eqref{H:t2} vanishes for the $v^\mu$-projection and the resulting expression 
becomes explicitly translation invariant (up to twist-four corrections) 
\begin{align}
\langle p^\prime|  [\mathcal{O}^{(\slashed{v})}(z_1+\delta,z_2+\delta)]^{t2}|p\rangle  &=
e^{-2i\delta\xi(Pv)} \langle p^\prime|  [\mathcal{O}^{(\slashed{v})}(z_1,z_2)]^{t2}|p\rangle
=
e^{i\delta(\Delta v)} \langle p^\prime|  [\mathcal{O}^{(\slashed{v})}(z_1,z_2)]^{t2}|p\rangle
\end{align}
provided the skewedness parameter $\xi$ is defined with respect to the $v^\mu$ vector, as in Eq.~\eqref{notations1}.
The full expression for an open Lorentz index does not transform properly, however:
\begin{align}
&\langle p^\prime|  [\mathcal{O}^{(\gamma_\mu)}(z_1+\delta,z_2+\delta)]^{t2}|p\rangle  =
  2 P_\mu e^{i\delta(\Delta v)}  \int_{-1}^1 dx\,  e^{- i (Pv) [z_1(\xi-x)+z_2(x+\xi)]} H(x,\xi) 
\notag\\&\quad
-  \Delta^\perp_\mu  \partial_\xi \Big\{ 
\int_0^1\!du\, e^{-2iu \xi (Pv)}\int_{-1}^1\! dx\,   e^{- iu (Pv) [z_1(\xi-x)+z_2(x+\xi)]}  H(x,\xi)\Big\}
\notag\\ &
\hspace*{5cm}
\slashed{=} e^{i\delta(\Delta v)} \langle p^\prime| [\mathcal{O}^{(\gamma_\mu)}(z_1,z_2)]^{t2}|p\rangle \,,
\end{align}
and this pathology has to be be cured by adding the twist-three contributions. 

Note that for a spin-zero target 
\begin{align}
 \langle p^\prime|  [\mathcal{O}^{(\slashed{v}\gamma_5)}(z_1,z_2)]^{t2}|p\rangle  &= 0\,,
\end{align}
so that the general expressions in \eqref{t3} are simplified to
\begin{align}
&[\mathcal{O}^{(\gamma_\mu)}(z_1,z_2)]^{t3} =
\frac12 [(vd)\partial_\mu - (v\partial) d_\mu] \int_0^1\!\!udu \!\int_{z_2}^{z_1}\frac{dw}{z_{12}} 
\Big[z_1 \mathcal{O}^{(\slashed{v})}(uz_1,uw) + z_2 \mathcal{O}^{(\slashed{v})}(uw,uz_2) \Big],
\notag\\
&[\mathcal{O}^{(\gamma_\mu\gamma_5)}(z_1,z_2)]^{t3} =
-\frac{i}{2} \epsilon^{\rho\nu\sigma\mu} 
v_\rho \partial_\sigma d_\nu  \int_0^1\!udu \!\int_{z_2}^{z_1}\frac{dw}{z_{12}} \Big [z_1 \mathcal{O}^{(\slashed{v})}(uz_1,uw)
 - z_2 \mathcal{O}^{(\slashed{v})}(uw,uz_2)\Big].
\label{t3-scalar}
\end{align} 
We will continue with the vector operator as the more relevant one. Taking the matrix element $\langle p'|\ldots |p\rangle$ one obtains 
after a little algebra
\begin{align}
\langle p^\prime|  [\mathcal{O}^{(\gamma_\mu)}(z_1,z_2)]^{t3}|p\rangle  &=  
 i (vP) \Delta_\mu ^\perp  \int_0^1\!udu \!\int_{z_2}^{z_1}\frac{dw}{z_{12}} 
\int_{-1}^1 \!\!dx\, 
\Big[ z_1 e^{- i u  \ell_{z_1w} }  + z_2  e^{- i u \ell_{wz_2}} \Big]
\mathcal{D}H(x,\xi)\,, 
\label{H:t3}
\end{align}
where
\begin{align}
   \ell_{z_1z_2} & = [z_1(\xi-x)+z_2(x+\xi)] (Pv)   =  - z_{12} x (Pv) - \frac12(z_1+z_2)(\Delta v)  
\label{L}
\end{align}
and 
\begin{align}
  \mathcal{D} = x\partial_x+\xi\partial_\xi\,.
\label{D}
\end{align}
The contribution in the second line in \eqref{H:t2} can be rewritten as 
(we suppress an overall factor $ \Delta^\perp_\mu$)
\begin{align}
-  \partial_\xi \int_0^1\!du \int_{-1}^1\!\! dx\,e^{- i u \ell_{z_1z_2}}  H(x,\xi) =
\notag\\
&\hspace*{-3cm}= - \frac{1}{\xi} \int_{-1}^1\!\! dx\,  e^{- i \ell_{z_1z_2}} H(x,\xi) 
- \frac{1}{\xi}\int_0^1\!du \int_{-1}^1\!\! dx\, e^{- i u \ell_{z_1z_2}}  \mathcal{D}  H(x,\xi)\,.
\end{align}
The first term on the r.h.s. of this expression has the structure consistent with translation invariance, and the 
second, noninvariant, term can be combined with the twist-three contribution in Eq.~\eqref{H:t3}.  The resulting expression can be brought to the form
\begin{align}
&\int_0^1\!du\, \int_{-1}^1\!\! dx\,
\biggl\{ - \frac{1}{\xi} e^{- i u \ell_{z_1z_2}} 
+
 i u (vP)\!\int_{z_2}^{z_1}\frac{dw}{z_{12}}  
\Big[z_1 e^{- i u  \ell_{z_1w} }  + z_2  e^{- i u \ell_{wz_2}} \Big]
\biggr\}
\mathcal{D}H(x,\xi) 
\notag\\&
= \int_{-1}^1\!\!dx\, e^{- i  \ell_{z_1z_2}} \mathbb{H}(x,\xi)
\label{mathbbH}
\end{align}
with 
\begin{align}
 & \mathbb H(x,\xi)|_{x>\xi} = 
- \frac{1}{2\xi} \int_{x}^1 \!\! dy\,
\biggl(\frac{1}{y-\xi}  + \frac{1}{y+\xi}  \biggr) \mathcal{D} H(y,\xi)\,, 
\notag\\
& \mathbb H(x,\xi)|_{-\xi<x<\xi} =
\frac{1}{2\xi} \int_{-1}^x \!\! dy\,
\frac{1}{y-\xi} \mathcal{D}H(y,\xi) - \frac{1}{2\xi} \int_{x}^1 \!\! dy\, \frac{1}{y+\xi}\mathcal{D}H(y,\xi)\,, 
\notag\\
& \mathbb H(x,\xi)|_{x<-\xi} =
 \frac{1}{2\xi} \int_{-1}^x \!\! dy\,
\biggl(
\frac{1}{y-\xi} + \frac{1}{y+\xi} \biggr) \mathcal{D} H(y,\xi)\,. 
\label{result-mathbbH}
\end{align}
The derivation is explained in Appendix~\ref{App:mathbbH}.

Finally, collecting all terms we obtain for the sum of twist-two and twist-three contributions
\begin{align}
&\langle p'|[\mathcal{O}^{(\gamma_\mu)}(z_1,z_2)]^{t2+t3}|p\rangle =
\notag\\
&=
\int_{-1}^1\!\!\!dx\,  e^{- i (Pv) [z_1(\xi-x)+z_2(x+\xi)]}
\biggl\{
2 P_\mu H(x,\xi) +
 \Delta^\perp_\mu \biggl[\mathbb H(x,\xi)\! -\!\frac{1}{\xi} H(x,\xi)\biggr]
\biggr\},
\label{H:t23}
\end{align}
so that the translation invariance is indeed restored:
\begin{align}
 \langle p'|[\mathcal{O}^{(\gamma_\mu)}(z_1+\delta,z_2+\delta)]^{t2+t3}|p\rangle 
&=
e^{i\delta(\Delta v)} \langle p^\prime| [\mathcal{O}^{(\gamma_\mu)}(z_1,z_2)]^{t2+t3}|p\rangle
+ \mathcal{O}(v^\mu, v^2)\,. 
\end{align}
The combination 
\begin{align}
 G(x,\xi) &= \mathbb H(x,\xi) -\frac{1}{\xi} H(x,\xi)
\end{align}
can be viewed as the twist-three pseudo- or quasi-GPD, which coincide to the present accuracy. 
It can be rewritten as shown in Eq.~\eqref{result:scalar} to make explicit that there is no  $1/\xi$ enhancement.

The axial-vector contribution in \eqref{t3-scalar} is calculated in a similar manner. 
One obtains
\begin{align}
 &\langle p'|[\mathcal{O}^{(\gamma_\mu\gamma_5)}(z_1,z_2)]^{t3}|p\rangle =
\notag\\&=
 - \epsilon^{\rho\nu\sigma\mu} v_\rho\Delta_\nu P_\sigma \int_0^1\!udu \!\int_{z_2}^{z_1}\frac{dw}{z_{12}} 
 \int_{-1}^1 \!\!dx\,
\Big[ 
 z_1 e^{- i u  \ell_{z_1w}} - z_2  e^{- i u  \ell_{w z_2}} \Big]
\mathcal{D} H(x,\xi)\,.
\end{align}  
This expression can be rewritten as
\begin{align}
\langle p'|[\mathcal{O}^{(\gamma_\mu\gamma_5)}(z_1,z_2)]^{t3}|p\rangle &=
  \widetilde{\Delta}_\mu^\perp \int_{-1}^1\!\!dx\,  e^{- i (Pv) [z_1(\xi-x)+z_2(x+\xi)]} \widetilde{G}(x,\xi)\,,
\end{align}
where
\begin{align}
&\widetilde{G}(x,\xi)|_{x>\xi} = 
- \frac{1}{2\xi} \int_{x}^1 \!\! dy\,
\biggl(\frac{1}{y-\xi}  - \frac{1}{y+\xi} \biggr)\mathcal{D} H(y,\xi) 
\notag\\
& \widetilde{G}(x,\xi)|_{-\xi<x<\xi} =
\frac{1}{2\xi} \int_{-1}^x \!\! dy\,
\frac{1}{y-\xi}  \mathcal{D} H(y,\xi) 
+ \frac{1}{2\xi} \int_{x}^1 \!\! dy\,
\frac{1}{y+\xi} \mathcal{D} H(y,\xi)\,, 
\notag\\
&\widetilde{G}(x,\xi)|_{x<-\xi} =
 \frac{1}{2\xi} \int_{-1}^x \!\! dy\,
\biggl(\frac{1}{y-\xi} - \frac{1}{y+\xi} \biggr) \mathcal{D} H(y,\xi)\,,
\end{align}
and
\begin{align}
 \epsilon^\perp_{\mu\nu} = \epsilon_{\mu\nu\alpha\beta} \frac{P_\alpha v_\beta}{(Pv)}\,,
\qquad 
\widetilde{\Delta}^\perp_{\mu} = i  \epsilon^\perp_{\mu\nu} \Delta^\nu.
\end{align}
Our results agree with Ref.~\cite{Belitsky:2000vx} where the derivation was done for symmetric points and 
using a different method.

\section{Nucleon target}

The calculation for spin-1/2 targets follows the same routine but is more cumbersome due to proliferation 
of Dirac structures.  Let
\begin{align}
& h_\rho = \bar u(p') \gamma_\rho u(p)\,, 
&& \tilde h_\rho = \bar u(p') \gamma_\rho\gamma_5 u(p)\,, 
\notag\\
&e_\rho = \bar u(p')\frac{i\sigma^{\rho\alpha}\Delta_\alpha}{2m} u(p)\,,
&& \tilde e_\rho = \frac{\Delta_\rho}{2m} \bar u(p') \gamma_5 u(p)\,. 
\label{Dirac}
\end{align}
Nucleon GPDs are defined~\cite{Diehl:2003ny,Belitsky:2005qn} as matrix elements of (renormalized)
quark-antiquark operators with light-like separations, $n^2=0$,  
\begin{align}\label{defNucleonGPD1}
\langle p^\prime|\mathcal {O}^{(\slashed{n})}(z_1,z_2)|p\rangle &= 
\int_{-1}^1\!\! dx\, e^{-i (Pn)[z_1(\xi-x)+z_2(x+\xi)]}
\Big\{ (n h)  H(x,\xi) + (n e) E(x,\xi)\Big\},  
\notag\\
\langle p^\prime|\mathcal {O}^{(\slashed{n}\gamma_5)}(z_1,z_2)|p\rangle &= 
\int_{-1}^1\!\! dx\, e^{-i (Pn)[z_1(\xi-x)+z_2(x+\xi)]}
\Big\{ (n \tilde h)  \widetilde H(x,\xi) + (n \tilde e) \widetilde E(x,\xi) \Big\}.  
\end{align}
At tree level and to twist-two accuracy, the same expressions are valid for the pseudo- and quasi-GPDs, with
the replacement $n_\mu\mapsto v_\mu$. The main difference to the spin-zero case is that the leading-twist 
axial-vector contributions do not vanish.  Thus complete expressions for the twist-three contributions at the 
operator level \eqref{t3} have to be used. In the following equations we will use a 
shorthand notation for the scalar products  
$
   P_v = (Pv)
$,
etc.

\subsection{Vector operator}

Following Ref.~\cite{Belitsky:2000vx} we write the result in the form
\begin{align}
 \langle p^\prime|[\mathcal {O}^{(\gamma_\mu)}(z_1,z_2)]^{t2+t3}|p\rangle &= 
 \int_{-1}^1\!dx\, e^{- i P_v [z_1(\xi-x)+z_2(x+\xi)]}
\biggl\{
2 P_\mu V_1 (x,\xi) +
\notag\\&\quad +
\Delta^\perp_\mu \biggl[\mathbb{V}_1(x,\xi) -\frac{1}{\xi} V_1(x,\xi)\biggr]
+ \left( h_\mu - \frac{P_\mu}{P_v} h_v \right) \mathbb{V}_2(x,\xi)
\notag\\&\quad
+ \widetilde{\Delta}_\mu^\perp \mathbb{V}_3(x,\xi)
+  i\epsilon_{\mu\rho}^\perp \left(\tilde h_\rho - \frac{P_\rho}{P_v} \tilde h_v \right) \mathbb{V}_4(x,\xi)
\biggr\},
\end{align}
where
\begin{align}
 V_1(x,\xi) &= \frac{1}{2 P_v} \Big[ h_v H(x,\xi) + e_v E(x,\xi)\Big]\,, 
&&
 V_2(x,\xi) = H(x,\xi) + E(x,\xi)\,,
\notag\\
 V_3(x,\xi) &= \frac{1}{2P_v}\Big[\tilde h_v \widetilde H(x,\xi) + \tilde e_v \widetilde E(x,\xi)\Big]\,,
&&
V_4(x,\xi) = \widetilde{H}(x,\xi)\,,
\end{align}
and the functions $\mathbb{V}_{1-4}$ are given by the following expressions~\cite{Belitsky:2000vx}:
\begin{align}
&\mathbb{V}_1(x,\xi)|_{x>\xi}= 
- \frac{1}{2\xi} \int_{x}^1 \!\! dy\,
\biggl(\frac{1}{y-\xi}  + \frac{1}{y+\xi} \biggr) \mathcal{D} V_1(y,\xi)\,, 
\notag\\
&\mathbb{V}_1(x,\xi)|_{-\xi<x<\xi} =
\frac{1}{2\xi} \int_{-1}^x \!\! dy\, \frac{1}{y-\xi} \mathcal{D} V_1(y,\xi) 
- \frac{1}{2\xi} \int_{x}^1 \!\! dy\, \frac{1}{y+\xi}\mathcal{D} V_1(y,\xi)\,, 
\notag\\
&\mathbb{V}_1(x,\xi)|_{x<-\xi}=
 \frac{1}{2\xi} \int_{-1}^x \!\! dy\,
\biggl(\frac{1}{y-\xi} + \frac{1}{y+\xi} \biggr) \mathcal{D} V_1(y,\xi)\,. 
\end{align}
\begin{align}
 & \mathbb V_2(x,\xi)|_{x>\xi} = 
\phantom{+} \frac{1}{2} \int_{x}^1 \!\! dy\,
\biggl(\frac{1}{y-\xi}  + \frac{1}{y+\xi} \biggr)  V_2 (y,\xi)\,, 
\notag\\
& \mathbb V_2(x,\xi)|_{-\xi<x<\xi} =
-\frac{1}{2} \int_{-1}^x \!\! dy\,\frac{1}{y-\xi} V_2(y,\xi) 
+ \frac{1}{2} \int_{x}^1 \!\! dy\,\frac{1}{y+\xi} V_2(y,\xi)\,, 
\notag\\
& \mathbb V_2(x,\xi)|_{x<-\xi} =
 - \frac{1}{2} \int_{-1}^x \!\! dy\,
\biggl(\frac{1}{y-\xi} + \frac{1}{y+\xi} \biggr) V_2(y,\xi)\,, 
\end{align}
\begin{align}
&\mathbb{V}_3(x,\xi)|_{x>\xi} = 
- \frac{1}{2\xi} \int_{x}^1 \!\! dy\,
\biggl(
\frac{1}{y-\xi}  - \frac{1}{y+\xi}  
\biggr) \mathcal{D} V_3(y,\xi)\,, 
\notag\\
&\mathbb{V}_3(x,\xi)|_{-\xi<x<\xi} =
\frac{1}{2\xi} \int_{-1}^x \!\! dy\,\frac{1}{y-\xi} \mathcal{D} V_3(y,\xi) 
+ \frac{1}{2\xi} \int_{x}^1 \!\! dy\,\frac{1}{y+\xi}  \mathcal{D} V_3(y,\xi)\,, 
\notag\\
&\mathbb{V}_3(x,\xi)|_{x<-\xi} =
 \frac{1}{2\xi} \int_{-1}^x \!\! dy\,
\biggl(
\frac{1}{y-\xi} - \frac{1}{y+\xi} \biggr) \mathcal{D} V_3(y,\xi)\,, 
\end{align}
\begin{align}
& \mathbb{V}_4(x,\xi)|_{x>\xi} = 
\phantom{+}\frac{1}{2} \int_{x}^1 \!\! dy\,
\biggl(\frac{1}{y-\xi}  - \frac{1}{y+\xi}  \biggr) \widetilde H(y,\xi)\,, 
\notag\\
&\mathbb{V}_4(x,\xi)|_{-\xi<x<\xi} =
-\frac{1}{2} \int_{-1}^x \!\! dy\,\frac{1}{y-\xi} \widetilde H(y,\xi) 
- \frac{1}{2} \int_{x}^1 \!\! dy\,\frac{1}{y+\xi} \widetilde H(y,\xi)\,, 
\notag\\
& \mathbb{V}_4(x,\xi)|_{x<-\xi} =
 - \frac{1}{2} \int_{-1}^x \!\! dy\,
\biggl(\frac{1}{y-\xi} - \frac{1}{y+\xi} \biggr) \widetilde H(y,\xi)\,. 
\end{align}
It is tacitly assumed that the differential operator $\mathcal{D}$ \eqref{D} only acts on the 
GPDs, but not on the Dirac/Lorentz structures, i.e.
\begin{align}
 \mathcal{D} V_1(x,\xi) &\equiv 
\frac{1}{2 P_v} \Big[ h_v \mathcal{D} H(x,\xi)+ e_v \mathcal{D} E(x,\xi)\Big]. 
\end{align}
Note also that 
$
 h_\mu - \frac{P_\mu}{P_v} h_v  = e_\mu - \frac{P_\mu}{P_v} e_v  
$,
so that $\mathbb V_2$ can be written in a more symmetric form involving
$$
 \left( h_\mu - \frac{P_\mu}{P_v} h_v \right) H(x,\xi) + \left(e_\mu - \frac{P_\mu}{P_v} e_v\right)  E(x,\xi)\,.
$$

\subsection{Axial-vector operator}

The results for the axial-vector operator are very similar and can be obtained by
the substitution 
\begin{align}
 h_v H + e_v E &~\longleftrightarrow~ \tilde h_v\widetilde H +  \tilde e_v \widetilde E\,,
\notag\\
 \Big( h_\mu - \frac{P_\mu}{P_v} h_v \Big) (H + E) &~\longleftrightarrow~
 \Big( \tilde h_\mu - \frac{P_\mu}{P_v} \tilde h_v \Big) \widetilde H\,.
\end{align}
%
%
One gets
\begin{align}
 \langle p^\prime|[\mathcal {O}^{(\gamma_\mu\gamma_5)}(z_1,z_2)]^{t2+t3}|p\rangle &= 
 \int_{-1}^1\!dx\, e^{- i P_v [z_1(\xi-x)+z_2(x+\xi)]}
\biggl\{
2 P_\mu A_1 (x,\xi) +
\notag\\&\quad +
\Delta^\perp_\mu \biggl[\mathbb{A}_1(x,\xi) -\frac{1}{\xi} A_1(x,\xi)\biggr]
+ \left( \tilde h_\mu - \frac{P_\mu}{P_v} \tilde h_v \right) \mathbb{A}_2(x,\xi)
\notag\\&\quad
+ \widetilde{\Delta}_\mu^\perp \mathbb{A}_3(x,\xi)
+  i\epsilon_{\mu\rho}^\perp \left( h_\rho - \frac{P_\rho}{P_v} h_v \right) \mathbb{A}_4(x,\xi)
\biggr\},
\end{align}
where
\begin{align}
  A_1(x,\xi) &=  \frac{1}{2 P_v} \Big[ \tilde h_v \widetilde H(x,\xi) + \tilde e_v \widetilde E(x,\xi)\Big], 
&&
A_2(x,\xi) = \widetilde{H}(x,\xi)\,,
\notag\\
 A_3(x,\xi) &= \frac{1}{2P_v}\Big[h_v H(x,\xi) + e_v  E(x,\xi)\Big], 
&&
A_4(x,\xi) = H(x,\xi)+ E(x,\xi)\,,
\end{align}
and the functions $\mathbb{A}_{1-4}$ are given in terms of $A_{1-4}$ by the same expressions as in the vector case:
\begin{align}
&\mathbb{A}_1(x,\xi)|_{x>\xi}= 
- \frac{1}{2\xi} \int_{x}^1 \!\! dy\,
\biggl(\frac{1}{y-\xi}  + \frac{1}{y+\xi} \biggr) \mathcal{D} A_1(y,\xi)\,, 
\notag\\
&\mathbb{A}_1(x,\xi)|_{-\xi<x<\xi} =
\frac{1}{2\xi} \int_{-1}^x \!\! dy\, \frac{1}{y-\xi} \mathcal{D} A_1(y,\xi) 
- \frac{1}{2\xi} \int_{x}^1 \!\! dy\, \frac{1}{y+\xi}\mathcal{D} A_1(y,\xi)\,, 
\notag\\
&\mathbb{A}_1(x,\xi)|_{x<-\xi}=
 \frac{1}{2\xi} \int_{-1}^x \!\! dy\,
\biggl(\frac{1}{y-\xi} + \frac{1}{y+\xi} \biggr) \mathcal{D} A_1(y,\xi)\,,
\end{align}
\begin{align}
 &\mathbb A_2(x,\xi)|_{x>\xi} = 
\phantom{+} \frac{1}{2} \int_{x}^1 \!\! dy\,
\biggl( \frac{1}{y-\xi}  + \frac{1}{y+\xi}  \biggr) A_2 (y,\xi)\,, 
\notag\\
 & \mathbb A_2(x,\xi)|_{x<-\xi} =
 - \frac{1}{2} \int_{-1}^x \!\! dy\,
\biggl(\frac{1}{y-\xi} + \frac{1}{y+\xi} \biggr) A_2(y,\xi)\,, 
\notag\\
 & \mathbb A_2(x,\xi)|_{-\xi<x<\xi} =
-\frac{1}{2} \int_{-1}^x \!\! dy\,\frac{1}{y-\xi} A_2(y,\xi) 
+ \frac{1}{2} \int_{x}^1 \!\! dy\,\frac{1}{y+\xi} A_2(y,\xi)\,, 
\end{align}
\begin{align}
&\mathbb{A}_3(x,\xi)|_{x>\xi} = 
- \frac{1}{2\xi} \int_{x}^1 \!\! dy\,
\biggl(\frac{1}{y-\xi}  - \frac{1}{y+\xi}  \biggr) \mathcal{D} A_3(y,\xi)\,, 
\notag\\
&\mathbb{A}_3(x,\xi)|_{-\xi<x<\xi} =
\frac{1}{2\xi} \int_{-1}^x \!\! dy\,\frac{1}{y-\xi} \mathcal{D} A_3(y,\xi) 
+ \frac{1}{2\xi} \int_{x}^1 \!\! dy\,\frac{1}{y+\xi}  \mathcal{D} A_3(y,\xi)\,, 
\notag\\
&\mathbb{A}_3(x,\xi)|_{x<-\xi} =
 \frac{1}{2\xi} \int_{-1}^x \!\! dy\,
\biggl(\frac{1}{y-\xi} - \frac{1}{y+\xi} \biggr) \mathcal{D} A_3(y,\xi)\,, 
\end{align}
\begin{align}
& {\mathbb A}_4(x,\xi)|_{x>\xi} = 
\phantom{+} \frac{1}{2} \int_{x}^1 \!\! dy\,
\biggl(\frac{1}{y-\xi}  - \frac{1}{y+\xi}\biggr)   A_4(y,\xi)\,, 
\notag\\
& {\mathbb A}_4(x,\xi)|_{-\xi<x<\xi} =
-\frac{1}{2} \int_{-1}^x \!\! dy\,\frac{1}{y-\xi}  A_4(y,\xi) 
- \frac{1}{2} \int_{x}^1 \!\! dy\,\frac{1}{y+\xi}  A_4(y,\xi)\,, 
\notag\\
& {\mathbb A}_4(x,\xi)|_{x<-\xi} =
 - \frac{1}{2} \int_{-1}^x \!\! dy\,
\biggl(
\frac{1}{y-\xi} - \frac{1}{y+\xi} \biggr)  A_4(y,\xi)\,. 
\end{align}

\section{Numerical Estimates}

\begin{figure}[t]
\centerline{\includegraphics[width=0.66\textwidth]{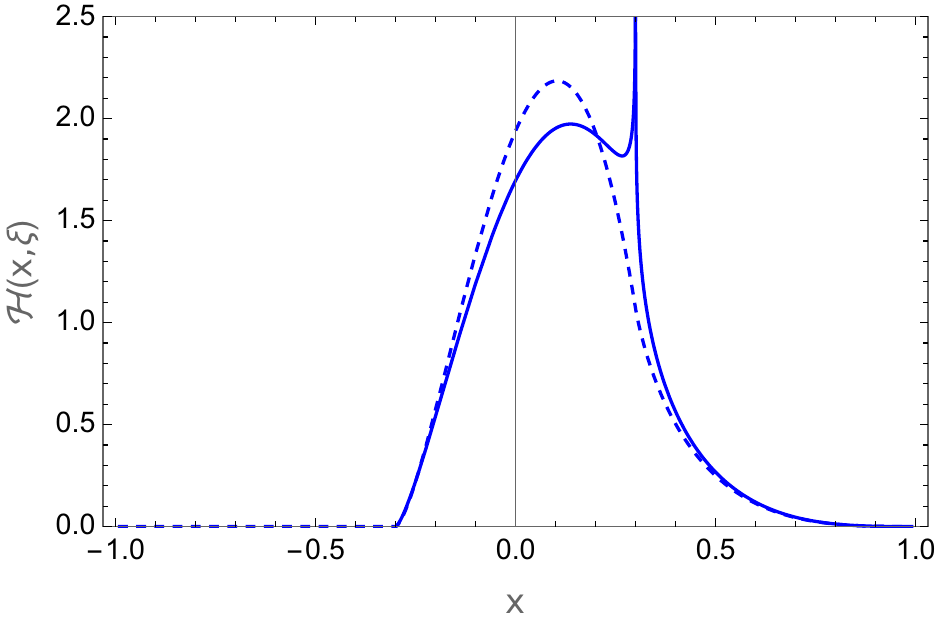}}
\caption{The quasi-GPD (pseudo-GPD) $\mathcal{H}(x,\xi)$ \eqref{qGPD} for $\xi=0.3$ including kinematic 
twist-three contributions (solid line). The GPD model \eqref{Hmodel} is shown by dashes for comparison.}
\label{fig:WWqGPD}
\end{figure}

As an illustration of  possible size of the kinematic twist-three contribution we 
assume that the calculation is done for the time projection $\gamma_\mu\mapsto \gamma_0$ for the scalar 
target~\eqref{result:scalar}. The resulting quasi- (alias pseudo-) PDF including twist-three corrections 
\begin{align}
    \mathcal{H}(x,\xi) = H(x,\xi) + \frac{\Delta_0^\perp}{2P_0} G(x,\xi)  
\label{qGPD}
\end{align}
is shown in Figure \ref{fig:WWqGPD} assuming an {\it ad hoc} value
\begin{align}
  \frac{\Delta_0^\perp}{2P_0} = 0.1\,.    
\end{align}  
In this calculation  we used a simple valence GPD model~\cite{Belitsky:2005qn}.
\begin{align}
H(x,\xi) & =  \theta(x>-\xi) \frac{1-n/4}{\xi^3} \left(\frac{x+\xi}{1+\xi}\right)^{2-n} [\xi^2 -x + (2-n) \xi (1-x)]
\notag\\&\quad
- \theta(x>\xi) \frac{1-n/4}{\xi^3} \left(\frac{x-\xi}{1-\xi}\right)^{2-n} [\xi^2 -x - (2-n) \xi (1-x)]
\label{Hmodel}
\end{align}
with $n=1/2$ corresponding to the Bjorken-x dependence of the valence quark PDF $q_v(x)\sim x^{-1/2}(1-x)^3$.  
For this plot we have chosen $\xi =0.3$. The GPD model \eqref{Hmodel} used as an input is shown by dashes for comparison.

One sees that the twist-three contributions are generally quite sizable and strongly enhanced at $x\to \xi$ because 
of the logarithmic divergence $\sim\ln(x-\xi)$. It is interesting to note that the twist-three contribution is discontinuous at $x=\xi$
and the difference between the values extrapolated from the DGLAP and ERBL regions is finite, see Eq.~\eqref{disc}.
We have to caution that this plot is provided only for the illustration; the real size of the kinematic correction
will depend strongly on the kinematics of a particular lattice calculation. This is especially true for the 
nucleon target due to multitude of the various contributions. Note that contributions of 
all four existing nucleon GPDs are intertwined at twist-three level, so that the magnitude of the correction 
will depend on the relative size of the GPDs $H$, $E$, etc., as well. 
A detailed investigation of these dependencies goes beyond the tasks of this work.

\section{Conclusions}

We have presented explicit expressions for the kinematic twist-three corrections to quasi- or pseudo-GPDs 
for spin-zero and spin-1/2 targets.
The results agree with Refs.~\cite{Belitsky:2000vx,Kivel:2000rb,Kivel:2000cn} where the derivation was done 
for symmetric points only and using a different method.
Apart from providing an independent check, a new contribution of this work is to consider general kinematics.
In this context, we demonstrate that kinematic corrections restore the proper properties of the
quasidistributions under translations along the quark-antiquark separation axis (up to twist-four terms). 
They have to be either taken into account explicitly in lattice calculations
of GPDs in the pseudo- or quasi-PDF approach, or removed by taking suitable projections in Lorentz/Dirac indices. 
Otherwise, the difference in the results obtained using symmetric and asymmetric frames would need to be considered 
as an intrinsic uncertainty of the calculation.

The ``genuine'' twist-three contributions related to quark-gluon correlations induce
violation of translation invariance at twist-four level only and are not 
included in our expressions. They can be deduced from Ref.~\cite{Belitsky:2000vx} (in symmetric kinematics). 
Such extra terms involve new and complicated nonperturbative input and can be added
when lattice calculations have sufficient accuracy. 

An extension of our results to kinematic twist-four contributions presents a natural next task.
For moments of quasidistributions, such corrections can be obtained from the results presented in 
Ref.~\cite{Braun:2011dg} by simple algebra. The problem is that kinematic twist-four corrections for the moments 
involve very nontrivial coefficients. In the application to DVCS, these complicated terms are canceled by the 
contributions from gluon emission from the hard quark propagator and the results can be resummed in a nonlocal 
expression in terms of the GPDs. The present case is more challenging and requires a dedicated study. 

\section*{Acknowledgments}
\addcontentsline{toc}{section}{Acknowledgments}

The author is grateful to A.~Manashov for discussions.
This work was supported in part by the Research Unit FOR2926 funded by the Deutsche
Forschungsgemeinschaft (DFG, German Research Foundation) under grant 409651613.


\appendix
\addcontentsline{toc}{section}{Appendices}
\renewcommand{\theequation}{\Alph{section}.\arabic{equation}}
\renewcommand{\thesection}{{\Alph{section}}}
\renewcommand{\thetable}{\Alph{table}}
\setcounter{section}{0} \setcounter{table}{0}
\section*{Appendices}

\section{Derivation of Eq.~\eqref{H:t2}}\label{App:t2}

Consider
\begin{align}
\langle p^\prime|  [\mathcal{O}^{(\gamma_\mu)}(z_1,z_2)]^{t2}|p\rangle  &=
\frac{\partial}{\partial v^\mu} 2 (Pv)\int_0^1\!du\! \int_{-1}^1\! dx\, e^{-i u \ell_{z_1z_2}}  H(x,\xi)\,,
\end{align}
where $\ell_{z_1z_2}$ is defined in \eqref{L}.
Note that the asymmetry parameter depends on $v^\mu$ as well,
\begin{align}
\frac{\partial\xi}{\partial v^\mu} & = -\frac{1}{2(Pv)} \Delta_\mu^\perp.
\end{align}
so that the derivative can be applied in three ways, to the $(Pv)$ prefactor, to the exponential $e^{-i u \ell_{z_1z_2}}$, and
to the GPD $ H(x,\xi)$.
Thus we get
\begin{align}
\langle p^\prime|  [\mathcal{O}^{(\gamma_\mu)}(z_1,z_2)]^{t2}|p\rangle  &=
2 P_\mu \!\int_0^1\!\!du\! \int_{-1}^1\!\!\! dx\, e^{-iu \ell_{z_1z_2}}  H(x,\xi)
- \Delta_\mu^\perp\!\! \int_0^1\!\!du\! \int_{-1}^1\!\!\! dx\,  e^{- iu \ell_{z_1z_2}}\partial_\xi H(x,\xi)
\notag\\&\quad 
+ 2 (P v) \int_0^1\!\!du\! \int_{-1}^1\!\! dx\,
(iu) \Big[z_{12} x P_\mu + \frac12(z_1\!+\!z_2) \Delta_\mu)\Big]
 e^{- iu \ell_{z_1z_2}}  H(x,\xi)\,. 
\end{align}
The last term can be simplified using that
\begin{align}
iu (Pv) \Big[z_{12} x P_\mu + \frac12(z_1+z_2) \Delta_\mu)\Big]  e^{-iu \ell_{z_1z_2}}
& = \Big[ P_\mu(x \partial_x  + \xi\partial_\xi) - \frac12 \Delta_\mu^\perp \partial_\xi\Big] e^{-iu \ell_{z_1z_2}}
\notag\\&=
\Big[ P_\mu u\partial_u - \frac12 \Delta_\mu^\perp \partial_\xi\Big] e^{-iu \ell_{z_1z_2}}
\end{align}
and integrating by parts over $u$ in the term $\sim P_\mu$. In this way one arrives at the expression in Eq.~\eqref{H:t2}.

\section{Derivation of Eq.~\eqref{result-mathbbH}}\label{App:mathbbH}

It is convenient to expand the exponential factors and write the expression of interest as a sum of moments
\begin{align}
\mathfrak{H}&= \int_0^1\!du\, \int_{-1}^1\!\! dx\,
\biggl\{ - \frac{1}{\xi} e^{- i u \ell_{z_1z_2}} 
+
 i u (vP)\!\int_{z_2}^{z_1}\frac{dw}{z_{12}}  
\Big[z_1 e^{- i u  \ell_{z_1w} }  + z_2  e^{- i u \ell_{wz_2}} \Big]
\biggr\}
\mathcal{D}H(x,\xi) 
\notag\\&
= \sum_N \frac{1}{N!} (-i(Pv))^N  \mathfrak{H}_N\,, 
\label{A1}
\end{align}
so that
\begin{align}
  \mathfrak{H}_N  &=  \int_{-1}^1\!\! dy\, [z_1(\xi-y)+z_2(y+\xi)]^N \mathbb{H}(y,\xi)\,,
\end{align}
where the function $ \mathbb{H}(y,\xi)$ is defined in \eqref{mathbbH}.
We obtain
\begin{align}
\mathfrak{H}_N&= -  \frac{1}{N+1} 
\int_{-1}^1 \!\!dx\, 
\biggl\{ \frac{z_1}{z_{12}} \frac{1}{x+\xi}
\Big\{ [2 z_1 \xi]^N -  [z_1 (\xi-x)+ z_2 (x+\xi)]^N\Big\}
\notag\\&\quad
+ \frac{z_2}{z_{12}} \frac{1}{x-\xi}\Big\{[2 z_2 \xi]^N - [z_1 (\xi-x)+z_2 (x+\xi)]^N\Big\}\biggr\}
\mathcal{D}H(x,\xi)
\notag\\&\quad
- \frac{1}{\xi} \frac{1}{N+1} \int_{-1}^1\!\! dx\, [z_1(\xi-x)+z_2(x+\xi)]^N\mathcal{D} H(x,\xi)
\notag\\&=
-  \frac{1}{N+1}\frac{1}{2\xi} \int_{-1}^1 \!\!dx\,
\sum\limits_{k=0}^{N-1}  [z_1 (\xi-x)+ z_2 (x+\xi)]^k
\biggl\{[2 z_1 \xi]^{N-k} + [2 z_2 \xi]^{N-k} \biggr\}
\mathcal{D}H(x,\xi)
\notag\\&\quad
- \frac{1}{\xi} \frac{1}{N+1} \int_{-1}^1\!\! dx\, [z_1(\xi-x)+z_2(x+\xi)]^N\mathcal{D} H(x,\xi)
\notag\\&=
-  \frac{1}{N+1}\frac{1}{2\xi} \int_{-1}^1 \!\!dx\,
\sum\limits_{k=0}^{N}  [z_1 (\xi-x)+ z_2 (x+\xi)]^k
\biggl\{[2 z_1 \xi]^{N-k} + [2 z_2 \xi]^{N-k} \biggr\}
\mathcal{D}H(x,\xi)\,.
\end{align}
Note that the role of the remnant of the twist-two contribution, the first term in Eq.~\eqref{A1}, 
is to extend the sum $\sum\limits_{k=0}^{N-1} \mapsto \sum\limits_{k=0}^{N}$. %
\footnote{For the axial-vector case there is no twist-two addition. In this case, the summation can be 
extended trivially because the two terms in curly brackets have opposite sign and cancel each other for $k=N$.} 
This sum can easily be taken so that one gets
\begin{align}
\mathfrak{H}_N&= 
- \frac{1}{2\xi} \frac{1}{N+1} \frac{1}{z_{12}}  
\int_{-1}^1 \!\!dx\,
\biggl\{\frac{1}{x+\xi}
\Big\{ [2 z_1 \xi]^{N+1} -  [z_1 (\xi-x)+ z_2 (x+\xi)]^{N+1}\Big\}
\notag\\&\quad
+ \frac{1}{x-\xi}\Big\{[2 z_2 \xi]^{N+1} - [z_1 (\xi-x)+z_2 (x+\xi)]^{N+1}\Big\}\biggr\}
\mathcal{D}H(x,\xi)\,.
\end{align}
Finally write
\begin{align}
\frac{1}{z_{12}} \frac{1}{N+1} \Big\{[2 z_2 \xi]^{N+1} - [z_1 (\xi-x)+z_2 (x+\xi)]^{N+1}\Big\} 
&=  \int_{\xi}^x\!\! dy\, [z_1(\xi-y)+z_2(y+\xi)]^N ,
\notag\\
\frac{1}{z_{12}} \frac{1}{N+1} \Big\{[2 z_1 \xi]^{N+1} - [z_1 (\xi-x)+z_2 (x+\xi)]^{N+1}\Big\} 
&=  \int_{-\xi}^x\!\! dy\, [z_1(\xi-y)+z_2(y+\xi)]^N ,
\end{align}
and interchange the order of integrations using
\begin{align}
  \int_{-1}^1 \!\!dx \int_\xi^x\!\!dy &=  
\int_{\xi}^1 \!\!dy \int_{y}^1 \!\!dx -  \int_{-1}^\xi \!\!dy \int_{-1}^y \!\!dx\,,  
\notag\\
  \int_{-1}^1 \!\!dx \int_{-\xi}^x\!\!dy &=  
   \int_{-\xi}^1 \!\!dy \int_{y}^1 \!\!dx -  \int_{-1}^{-\xi} \!\!dy \int_{-1}^y \!\!dx\,.  
\end{align}
In this way one gets
\begin{align}
\mathfrak{H}_N&=
\frac{1}{2\xi}\int_{-1}^1\!\! dy\, [z_1(\xi-y)+z_2(y+\xi)]^N 
\biggl\{- \theta(y-\xi) \int_{y}^1 \!\!\frac{dx}{x-\xi}
+ \theta(\xi-y) \int_{-1}^y \!\!\frac{dx}{x-\xi} 
\notag\\&\quad
- \theta(y+\xi) \int_{y}^1 \!\!\frac{dx}{x+\xi}
+  \theta(-\xi-y) \int_{-1}^y \!\!\frac{dx}{x+\xi} 
\biggr\} \mathcal{D} H(x,\xi)
\notag\\&= \int_{-1}^1\!\! dy\, [z_1(\xi-y)+z_2(y+\xi)]^N \mathbb{H}(y,\xi)\,,
\end{align}
where from the expression in \eqref{result-mathbbH} follows by a  simple reshuffling of terms.



\providecommand{\href}[2]{#2}\begingroup\raggedright\endgroup

\end{document}